\documentstyle[preprint,aps]{revtex}
\begin{document}
\draft
\title{IMPROVED PERTURBATION THEORY FOR \\
THE KARDAR-PARISI-ZHANG EQUATION}
\author{T. Blum and A. J. McKane}
\address{Department of Theoretical Physics \\
University of Manchester \\
Manchester M13 9PL, UK}
\date{\today}
\maketitle
\begin{abstract}
We apply a number of schemes which variationally improve perturbation
theory for the Kardar-Parisi-Zhang equation in order to extract estimates
for the dynamic exponent $z$.
The results for the various schemes show the same broad features,
giving closer agreement with numerical simulations in low dimensions
than self-consistent methods.
They do, however, continue to predict that $z=2$ in some critical
dimension $d_c$ in disagreement with the findings of simulations.
\end{abstract}
%\narrowtext
\section*{}
The Kardar-Parisi-Zhang (KPZ) equation \cite{KPZ} is perhaps the simplest
nonlinear stochastic equation of the diffusion type.
Nevertheless, the large-distance scaling behavior and anomalous
dimensions have not yet been understood on the basis of systematic
calculational schemes, such as the renormalization group.
To date the most reliable estimates for the anomalous dimensions,
obtained by direct analytic means from the KPZ equation, have come
from self-consistent or pseudo-variational procedures \cite{SE,BC,DMKB}.
These approaches are, in general, rather ad hoc and it would be useful
to have a method of improving estimates and perhaps gaining some idea of
their accuracy.
The relative merits of procedures such as these, genuine
variational procedures (by which we mean those with an associated bound)
and ``improvement" methods such as ``the principle of minimal sensitivity"
(PMS) \cite{PMS}, have recently been compared for stochastic processes
described by simple Langevin equations (or equivalently Fokker-Planck
equations) \cite{BM}.
In this Letter we extend these considerations to the KPZ equation ---
which can be formulated as a {\it functional} Fokker-Planck equation.
It turns out that, while genuine variational techniques are not
straightforward to use in this case, the application of the PMS yields
improved values for the dynamic exponent $z$ which agree better
with the results of numerical simulation of models believed
to be in the KPZ universality class.

The KPZ equation for surface growth in $(d+1)$ dimensions with random
deposition is
\begin{equation}
\dot h({\bf \vec x},t) \ = \
\nu_0 \nabla^2 h + g(\nabla h)^2 + \eta({\bf \vec x}, t),
\label{KPZ}
\end{equation}
where the single-valued function $h({\bf \vec x},t)$ represents the
interface and ${\bf \vec x}$ is a $d$-dimensional vector.
The subscript ``0" on surface tension $\nu_0$ and noise strength
$D_0$ (below) is used to distinguish these bare parameters from the
effective (renormalized) ones to be introduced later.
The noise $\eta({\bf \vec x},t)$ is Gaussian-distributed with zero
mean $\langle \eta({\bf \vec x},t) \rangle = 0$ and delta-function
correlations
\begin{equation}
\langle \eta({\bf \vec x}, t) \eta({\bf \vec x^{\prime}},t^{\prime})
\rangle \ = \ 2D_0 \delta({\bf \vec x} -{\bf \vec x^{\prime}})
\delta(t-t^{\prime}).
\label{noise-corr}
\end{equation}
Equivalently the noise may be specified in terms of the probability
density functional:
\begin{equation}
{\cal P}[\eta] \ \sim \
\exp \left\{ - (4D_0)^{-1} \int dt \int d^dx
\left[ \eta({\bf \vec x},t) \right]^2 \right\}.
\label{noise-dist}
\end{equation}
The KPZ equation (\ref{KPZ}) can then be taken as a transformation of
variables $\{ \eta({\bf \vec x},t) \rightarrow h({\bf \vec x},t)\}$
\cite{Graham} which leads to the following probability distribution of
the $h$ field:
\begin{equation}
{\cal P}[h] \ \sim \
\Biggl| {\rm Det} {\partial h({\bf \vec x},t) \over
\partial \eta({\bf \vec x^{\prime}},t^{\prime}) } \Biggr|
{\rm exp} \left\{ - {1 \over 4D_0} \int dt \int d^d x
\left[ \dot h -\nu_0\nabla^2 h -g(\nabla h)^2 \right]^2 \right\}.
\label{h-dist}
\end{equation}
The Jacobian above can be shown to be $h$-independent; it plays no role
in what follows and will be dropped.

Introducing the response field $\tilde h({\bf \vec x},t)$
\cite{Janssen} yields the distribution of the $h$ and $\tilde h$
fields as: ${\cal P}[h,\tilde h] \sim {\rm exp}
\bigl\{ {\cal S}_0[\tilde h, h]  \bigr\}$, where
\begin{equation}
{\cal S}_0[\tilde h, h] \ = \
\int dt \int d^dx \Bigl\{
D_0 \tilde h^2({\bf \vec x},t)
- \left[ \dot h -\nu_0\nabla^2 h -g(\nabla h)^2 \right]
\tilde h({\bf \vec x},t) \Bigr\}.
\label{hh-dist2}
\end{equation}
Note that the $\tilde h$'s are actually imaginary.
Fourier transforming the fields gives
\begin{eqnarray}
{\cal S}_0[\tilde h, h] \ = \
&&\int {d\omega \over 2 \pi} \int {d^d k \over (2\pi)^d}
\left[ D_0 \tilde h({\bf \vec k},\omega)\tilde h(-{\bf \vec k},-\omega)
-\left( \nu_0 {\bf \vec k}^2 -i\omega\right)
h({\bf \vec k},\omega)\tilde h(-{\bf \vec k},-\omega) \right]
\nonumber \\
-g &&\int {d\omega \over 2 \pi}
\int {d^d k \over (2\pi)^d}
\int {d\omega^{\prime} \over 2 \pi}
\int {d^d k^{\prime} \over (2\pi)^d}
[{\bf \vec k}\cdot {\bf \vec k^{\prime}}]
h({\bf \vec k},\omega)h({\bf \vec k^{\prime}},\omega^{\prime})
\tilde h(-{\bf \vec k}-{\bf \vec k^{\prime}},-\omega-\omega^{\prime}).
\label{hh-dist-ft}
\end{eqnarray}

The problem is completely solvable in the absence of the nonlinear
term ($g=0$) \cite{EW}, but one cannot access the strong-coupling fixed
point through perturbation and the renormalization group.
With these approaches ruled out, we turn to approximation techniques
such as variational methods and related procedures.
But just what kind of schemes are available?

Before answering this question, let us make the first step toward
a variational scheme by adding and subtracting an effective action
functional ${\cal S}_{\rm eff}[\tilde h, h]$
\begin{equation}
{\cal P}[h,\tilde h] \ \sim \
{\rm exp} \left\{ {\cal S}_{\rm eff}
+\lambda\left({\cal S}_0
-{\cal S}_{\rm eff} \right) \right\},
\label{hh-dist-pert}
\end{equation}
where $\lambda$ serves as a counting device and will eventually be
set to one.
The most satisfactory approach would be a genuine variational scheme.
Many of these are based on some version of the inequality
$e^{x} \ \geq \ 1 + x$ which can be extended to
\begin{equation}
e^{x} \ \geq \ \sum_{n=0}^{2N-1} {x^n \over n!}.
\label{bound}
\end{equation}
It is tempting to expand $\exp\{\lambda ({\cal S}_0
-{\cal S}_{\rm eff})\}$ and invoke this
inequality; however, it does not hold as the $\tilde h$'s are imaginary
and the above inequality applies to real $x$ only.

Since the existence of an imaginary part to ${\cal S}_0[\tilde h, h]$
appears to be the stumbling block, perhaps it would be better to
tackle the problem prior to introducing the response field.
The inequality would hold, but a new problem arises.
While one can proceed with a perturbation expansion around the bare
quadratic terms in (\ref{h-dist}), introducing variational quadratic
terms to serve as a basis for perturbation rearranges the series, and
one loses the property that divergences associated with the disconnected
diagrams cancel order-by-order.
These divergences render the inequality useless.

There are other genuine variational schemes based on the
differential form of the Fokker-Planck equation \cite{BM} and
applicable to problems with a finite number of degrees of freedom.
But they too are plagued with divergence problems when applied
to a field theory.
On the other hand, techniques are available which involve similar
computational steps to the variational one, although the philosophy is
rather different.
In particular, they do not involve bounds such as that given by
(\ref{bound}).
These techniques can be formulated by choosing a form for the
effective action ${\cal S}_{\rm eff}$ which is (i) quadratic (for
calculational convenience) and (ii) has the same structure as quadratic
terms in the bare action ${\cal S}_0$, but with $\nu_0q^2 \rightarrow
\nu_{\bf q}$ and $D_0 \rightarrow D_{\bf q}$, that is, with the bare
surface tension and bare noise strength replaced by their effective,
or renormalized, counterparts:
\begin{equation}
{\cal S}_{\rm eff}[\tilde h,h] \ = \
\int {d\omega \over 2 \pi} \int {d^d k \over (2\pi)^d}
\Bigl[ D_{\bf k} \tilde h({\bf \vec k},\omega)
\tilde h(-{\bf \vec k},-\omega)
-\left(\nu_{\bf k}-i\omega \right)
h({\bf \vec k},\omega)\tilde h(-{\bf \vec k},-\omega)\Bigr] .
\label{effective}
\end{equation}

The simplest scheme is probably the one pioneered by Edwards \cite{ES,SE},
where one chooses the functions $\nu _{\bf k}$ and $D_{\bf k}$
to be the ``exact" (but as yet unknown) surface tension and noise strength
respectively.
When ${\cal P}[h, \tilde{h}]$ is expanded in $\lambda$, the zeroth-order
term will give the correct result (by construction) and hence the rest of
the perturbation expansion (terms in $\lambda , \lambda ^2 , \ldots )$
will have to be identically zero.
This together with a scaling ansatz for the response and correlation
functions, give values for the exponent $z$ \cite{SE}, which we will
later compare to our findings.
This scheme has been called ``fastest apparent convergence" or FAC.

The approach based on the PMS is calculationally similar to those already
described --- it involves the same diagrams, since one is expanding the
same quantity about the same zeroth-order form (\ref{effective}), but
now the rationale is different.
The response or correlation function being calculated should not depend
on $\nu_{\bf q}$ or $D_{\bf q}$ , since these were introduced artificially
and are not part of the \lq\lq real" problem; however, any truncated
expansion does depend on them.
One then imposes a stationarity condition on the expansion to
determine the result which displays the least dependence on
them.
Thus one asks that the expansion mimic as best as possible one feature of
the true solution  --- its insensitivity to these variational terms ---
the hope being that other features will be mimicked as well.

As we have stressed, all the schemes discussed so far are based on a
perturbative calculation in the parameter $\lambda$.
The structure of the perturbation expansion and the diagrams is
essentially that given by Forster, Nelson and Stephen \cite{FNS}, except
that these authors were expanding in $g$ and we are expanding
in $\lambda$.
Thus we have quadratic, as well as cubic, terms in our ``interaction"
term and hence have terms odd in $\lambda$ in addition to the
more familiar $O(\lambda ^2 )$ terms of Ref. \cite{FNS}.

A perturbative calculation of $\langle |h({\bf \vec q},\omega)|^2\rangle$
to $O(\lambda^2)$ yields:
\begin{eqnarray}
\langle |h({\bf \vec q},\omega)|^2\rangle \ =\
&& {2 D_{\bf q} \over \left(\nu_{\bf q}^2+ \omega^2\right)}
+\lambda
\left[
{2 \left( D_0-D_{\bf q}\right) \over
\left(\nu_{\bf q}^2+ \omega^2\right)}
-{4 D_{\bf q} \nu_{\bf q}
\left(\nu_0q^2-\nu_{\bf q}\right)
\over \left(\nu_{\bf q}^2+ \omega^2\right)^2}
\right]
\nonumber \\
+\lambda^2 &&\left[
{2 D_{\bf q} \left(\nu_0q^2-\nu_{\bf q}\right)^2
\left(3\nu_{\bf q}^2-\omega^2\right)\over
\left(\nu_{\bf q}^2+ \omega^2\right)^3}
-{4 \left(D_0-D_{\bf q}\right) \nu_{\bf q}
\left(\nu_0q^2-\nu_{\bf q}\right)
\over \left(\nu_{\bf q}^2+ \omega^2\right)^2}
\right.
\nonumber \\
&& + {4g^2 \over \left(\nu_{\bf q}^2+ \omega^2\right)}
\int {d^d k \over (2 \pi)^d}
{ \left[{\bf \vec k}\cdot ({\bf \vec k+ \vec q})\right]^2
D_{\bf k}D_{\bf k+q} \left( \nu_{\bf k}+\nu_{\bf k+q} \right)
\over \nu_{\bf k}\nu_{\bf k+q}
\left(\left( \nu_{\bf k}+\nu_{\bf k+q} \right)^2 + \omega^2
\right)}
\nonumber \\
&&\left. - \left\{
{8g^2 D_{\bf q}\over \left(\nu_{\bf q}-i \omega\right)
\left(\nu_{\bf q}+ i\omega\right)^2}
\int {d^d k \over (2 \pi)^d}
{ \left[{\bf \vec k}\cdot {\bf \vec q}\right]
\left[{\bf \vec k}\cdot ({\bf \vec k+ \vec q})\right]
D_{\bf k} \over \nu_{\bf k}
\left(\nu_{\bf k}+\nu_{\bf k+q}  + i\omega \right)}
+c.c.\right\} \right],
\label{correlation}
\end{eqnarray}
where the internal frequency integral has been carried out.

There are various ways to proceed at this stage.
We want the calculation to be simple without sacrificing
any physics, so let us focus on the large-distance scaling
behavior.
Response and correlation functions calculated from (\ref{h-dist})
would be invariant under the scale transformation $\{{\bf \vec x}
\rightarrow b{\bf \vec x}, \ t \rightarrow b^z t, \  h \rightarrow
b^{\chi}h \}$ if the parameters scaled like $\{\nu \rightarrow b^{z-2}
\nu, D \rightarrow b^{-d-2\chi+z}D,g \rightarrow b^{z+\chi-2}g \}$.
We build these scalings into our effective surface tension
and noise strength by choosing
\begin{equation}
\nu_{\bf q} \ = \  Aq^z
\ \ \ {\rm and} \ \ \
D_{\bf q} \ = \  Bq^{-d-2\chi+z}.
\label{scaling}
\end{equation}
We also impose the exponent relation $z+\chi=2$ due to the Galilean
invariance \cite{FNS,HH}.
In what follows we will retain only the leading scaling behavior,
thus the bare terms $\nu_0$ and $D_0$ will no longer appear.

Two natural ways of simplifying Eq. (\ref{correlation}) are
first integrating over $\omega$ and second setting $\omega=0$.
These yield
\begin{equation}
\int{d \omega \over 2 \pi}\langle |h({\bf \vec q},\omega)|^2\rangle
= {B \over A}\bigl\{1+\lambda^2  \left[u
I_1(d,z) -2u I_2(d,z)\right] \bigr\}|{\bf \vec q}|^{-\Gamma},
\label{i-int}
\end{equation}
\begin{equation}
\langle |h({\bf \vec q},0)|^2  \rangle
={2B \over A^2} \bigl\{1+\lambda +\lambda^2\left[1  +u
J_1(d,z) - 4 u J_2(d,z) \right]\bigr\}|{\bf \vec q}|^{-z-\Gamma},
\label{j-int}
\end{equation}
respectively, where $\Gamma=d+4-2z$, $u=2g^2B/A^3$,
\begin{eqnarray}
I_1(d,z)= &&\int{d^d p \over (2 \pi)^d} { \left[{\bf \vec p}
\cdot ({\bf \vec p + \vec 1})\right]^2
|{\bf \vec p}|^{-\Gamma}|{\bf \vec p+ \vec 1}|^{-\Gamma} \over
\left(1+ |{\bf \vec p}|^{z}+|{\bf \vec p+ \vec 1}|^{z}\right)},
\label{i1} \\
I_2(d,z)= &&\int{d^d p \over (2 \pi)^d}
{ \left[{\bf \vec p} \cdot {\bf \vec 1}\right]
 \left[{\bf \vec p} \cdot ({\bf \vec p + \vec 1})\right]
|{\bf \vec p}|^{-\Gamma} \over
\left(1+ |{\bf \vec p}|^{z}+|{\bf \vec p+ \vec 1}|^{z}\right)},
\label{i2}
\end{eqnarray}
and $J_i(d,z)$ is the same as $I_i(d,z)$ (i=1,2), except that
the term in the denominator $ (1+|{\bf \vec p}|^z+
|{\bf \vec p+ \vec 1}|^{z})$ is replaced by
$(|{\bf \vec p}|^{z}+|{\bf \vec p+ \vec 1}|^{z})$.

When carrying out the PMS scheme we ask that the correlation functions
(\ref{i-int}) and (\ref{j-int}) be insensitive to the artificially
introduced quantities (\ref{scaling}).
Since $\nu_{\bf q}$ and $D_{\bf q}$ were assumed to have power-law
forms, they are, apart from the amplitudes $A$ and $B$, completely
characterized by the exponent $z$.
Thus, implementing the PMS in this case entails requiring that the
expressions (\ref{i-int}) and (\ref{j-int}) be independent of $z$.
There is, however, a subtlety here.
One cannot apply the PMS to a {\it function} and vary with respect
to a single parameter as that results in an overdetermination.
Either we perform a functional variation, or we apply the method
to a ${\bf \vec q}$-independent quantity.
Here we choose to vary with respect to $z$ alone, so we should
be analyzing a correlation function which has a ${\bf \vec q}$-independent
scaling form.
One possibility is to treat correlation functions of the type
$\langle (\nabla ^{\alpha} h)^2 \rangle$, which in ${\bf \vec q}$-space
reads $|{\bf \vec q}|^{2\alpha}\, \langle |h({\bf \vec q},\omega )|^2
\rangle$, rather than $\langle |h({\bf \vec q},\omega )|^2\rangle$.
Here we choose $\alpha$ so that the analogues of (\ref{i-int}) and
(\ref{j-int}) are ${\bf \vec q}$-independent.
Differentiating these expressions with respect to $z$ gives
\begin{eqnarray}
{\bf Scheme \ 1} && \hspace{1.5cm} \partial_z \left[ I_1(d,z)
- 2I_2(d,z) \right] = 0
\label{sch1} \\
{\bf Scheme \ 2} && \hspace{1.5cm}  \partial_z  \left[ J_1(d,z)
- 4J_2(d,z) \right] = 0,
\label{sch2}
\end{eqnarray}
respectively.

An alternative to starting from the full correlation function
$\langle | h({\bf \vec q},\omega )|^2 \rangle$, is to extract the
parts of this function which contribute to the renormalization of the
noise strength.
This is achieved by amputating the two external legs on the graphs for
the correlation function, that is, removing two factors of $G({\bf \vec q},
\omega )$ from (\ref{correlation}).
The resulting forms for the leading scaling behavior resemble
(\ref{i-int}) and (\ref{j-int}), but without the terms involving the
integrals $I_2$ and $J_2$.
Applying the PMS to the noise strength gives equations analogous
to (\ref{sch1}) and (\ref{sch2}):
\begin{eqnarray}
{\bf Scheme \ 3} && \hspace{1.5cm} \partial_z
\left[ I_1(d,z) \right] = 0
\label{sch3} \\
{\bf Scheme \ 4} && \hspace{1.5cm}  \partial_z
\left[ J_1(d,z) \right] = 0.
\label{sch4}
\end{eqnarray}

To find $z$ from (\ref{sch1})-(\ref{sch4}) we have to evaluate
integrals such as (\ref{i1}) and (\ref{i2}) for a range of values of $z$.
Going over to $d$-dimensional spherical polar coordinates, the
integrals may be reduced to double integrals depending on the parameters
$d$ and $z$.
We have taken great care in numerically integrating the double integrals
which give $\partial I_i /\partial z$ and $\partial J_i /\partial z$,
($i = 1,2$), since there are potential singularities at ${\bf \vec p} =
\vec 0 , -\vec 1$ and as ${\bf |\vec p |}\rightarrow \infty$.
In the first two cases the integrals converge if $z > 1$, which is always
true.
The singularities at large $|{\bf \vec p}|$ are more severe: \
integrals of the type (\ref{i1}) only exist if $z < (d + 4)/3$ and
those of the type (\ref{i2}) if $z \leq 2$.
In the latter case a naive analysis might suggest that the integral
only exists when $z < 1$, but the leading large $|{\bf \vec p}|$
term vanishes by ${\bf \vec p} \rightarrow  {\bf - \vec p}$ symmetry.

Based on these considerations, we have searched for solutions of
(\ref{sch1})-(\ref{sch4}) for parameter values $1 < z < (d + 4)/3$ for
$1 \leq d \le 2$ and $1 < z \leq 2$ for $d \ge 2$.
Within this parameter range there are integrable singularities which
numerical integration routines find difficult to cope with.
For this reason, we have used various transformations
which remove these integrable singularities, as well as different
routines, in order to check our computations.
Note that while schemes 3 and 4 seem to admit solutions with $z>2$,
it is easy to check that the subleading non-scaling terms which were
discarded to obtain (\ref{i-int}) and (\ref{j-int}) are only subdominant
if $z < 2$.
Thus the entire method is dependent on this bound being respected.

For Schemes 3 and 4, we readily find solutions, and the resulting
values of $z$ are reported in the table below.
We find no solutions to Schemes 1 and 2 as presented above.
However, the PMS methodology asserts that if the point of minimal
sensitivity  is not found by setting the derivative equal to zero, one
should next seek to minimize the derivative, and so on \cite{PMS}.
In this light we revise these schemes and look at higher derivatives.
Scheme 2 has a solution when the second derivative is set to zero
and the corresponding values of $z$ are reported in the table.
On the other hand, we have seen no evidence for a solution to Scheme
1 (in $d=2$) upon examining the second and third derivatives.

The simulation results included in the table are from the
hypercube-stacking model simulated in dimensions $d=1,2,3$
\cite{TFW} and the restricted solid-on-solid model, simulated up to
dimension $d=7$ \cite{AHKV}, both believed to belong to
the KPZ universality class.
The measurements of the exponent $\beta$, which governs the early
growth, agree between the models for $d=1,2,3$.
In the table we list the corresponding value for the dynamic
exponent $z=2/(\beta+1)$ and include the larger reported error bar.
The data for $d=4$ showed large fluctuations and finite-size
dependence, making it more difficult to establish accurate estimates
of the exponents \cite{AHKV}.

Schwartz and Edwards \cite{SE} adopted the FAC approach, determining $z$
by setting the terms of $O(\lambda)$ and higher in (\ref{i-int}) to
zero; the outcome of this procedure is included in the column labeled
SE.
The column labeled BC contains the findings of Bouchaud and Cates
\cite{BC} who pursued a self-consistent approach.
Actually the latter can also be formulated using the FAC criterion;
the only difference being that SE integrates over $\omega$ while BC
sets $\omega=0$.
Incidentally, the equations obtained by BC \cite{BC} can also be
derived as part of a functional PMS variation; but one in which scaling
is still ultimately put in by hand, and not deduced.

The values for $z$ furnished by the PMS Schemes (2-4) follow
the same general pattern: \ they fall below the exact value of $z=3/2$
in $d=1$ by a few percent; they then increase as $d$ increases and
eventually reach $z=2$ in some critical dimension $d_c$ (listed in the
table) in disagreement with the simulation results.
Note that the self-consistent methods contain enough of the correct
structure to yield the exact value of $z=3/2$ in $d=1$.
The PMS approach uses this same structure and information but in
a very different way and the exact result in $d=1$ is no longer
guaranteed.
However, for $d \geq 2$, the PMS schemes are more in line with the
simulation results.

In this Letter we have applied the PMS to obtain improved estimates for
the exponent $z$ in the KPZ equation.
The ideas discussed here could be extended in a number of ways.
Besides the obvious, but tedious, course of going to higher order
in the perturbation, one could adopt a more general (though still Gaussian)
form for (\ref{effective}) allowing for a frequency-dependent noise
strength $D_{{\bf k},\omega}$ or for a more general functional form
$G^{-1}({\bf \vec k},\omega)$ in place of the response
$ \nu _{\bf  k}-i\omega$.
This has proved successful within the self-consistent approach, where
a more sophisticated ansatz for the response function in Ref. \cite{DMKB}
led to improved results for $z$.
It would be interesting to see if more general ansatze such as these
would similarly improve on the estimates presented here.
A more ambitious calculation would be to apply the PMS to the
{\it function} (\ref{correlation}) without assuming scaling.
In the absence of a renormalization group treatment of this problem,
approaches such as these may be the most promising ways toward an
understanding of the anomalous large-distance behavior of the KPZ
equation.

\acknowledgements

We wish to thank J. P. Doherty for useful discussions and the Isaac
Newton Institute for Mathematical Sciences, where this work was begun,
for its hospitality.
TB acknowledges the support of the Engineering and Physical Sciences
Research Council under grant GR/H40150.

\newpage

\vbox{
\centerline{{\bf Table }}
\begin{displaymath}
\begin{tabular}{|c|c|c|c|c|c|c|}
\hline
\hline
\ \ $d$\ \  & Simulation & SE  & BC & Scheme 2 & Scheme 3 & Scheme 4\\
\tableline
1 & \ \ 1.500$\pm$0.001 \  \ & \ \ 1.500 \ \  & \ \ 1.500 \ \
& \ \ 1.424 \ \ & \ \ 1.420 \ \ & \ \  1.465 \ \  \\
2 & \ \ 1.613$\pm$0.003  \  \ & \ \ 1.705\ \  & \ \ 1.667 \ \
& \ \ 1.582 \ \ & \ \ 1.553 \ \ & \ \  1.621 \ \  \\
3 & \ \ 1.695$\pm$0.007 \  \ & \ \ 1.920 \ \  & \ \ 1.862 \ \
& \ \ 1.740 \ \ & \ \ 1.682 \ \ & \ \  1.769 \ \  \\
4 & \ \ 1.77$\pm$0.02 \  \ & \ \ ---  \ \  & \ \ ---  \ \
& \ \ 1.909 \ \ & \ \ 1.810 \ \ & \ \ 1.914 \ \  \\
\hline
$d_c$ & \ \ --- \  \ & \ \ 3.2 \ \  & \ \ 3.6 \ \
& \ \ 4.4 \ \ & \ \ 5.4 \ \ & \ \  4.6 \ \  \\
\hline
\hline
\end{tabular}
\end{displaymath}
\centerline{\small
Values of the dynamic exponent $z$.}
}

\end{document}